\documentclass[conference]{IEEEtran}
\IEEEoverridecommandlockouts
\usepackage{cite}
\usepackage{amsmath,amssymb,amsfonts}
\usepackage{algorithmic}
\usepackage{graphicx}
\usepackage{textcomp}
\usepackage{bbm}
\usepackage{xcolor}
\def\BibTeX{{\rm B\kern-.05em{\sc i\kern-.025em b}\kern-.08em
    T\kern-.1667em\lower.7ex\hbox{E}\kern-.125emX}}
    
\usepackage{tikz}
\usepackage{pgfplots}
\usetikzlibrary{spy}
\usepackage[normalem]{ulem}

\def\*#1{\mathbf{#1}}
\newcommand{\vect}[1]{{\mathbf{#1}}}
\newcommand{\mat}[1]{{\mathbf{#1}}}

\newcommand{\TM}{\mathcal{T}}
\newcommand{\cR}{\mathcal{R}}
\newcommand{\RM}{\mathcal{R}}
\renewcommand{\sf}[1]{\mathsf{#1}}

\newcommand{\cT}{\mathcal{T}}
\renewcommand{\intercal}{{T}}

\newcommand{\SDoF}{\textnormal{Sum-DoF}}
\newcommand{\SDoFlb}{\textnormal{Sum-DoF}_\textnormal{Lb}}

\newcommand{\K}{\mathsf{K}}
\renewcommand{\r}{\mathsf{r}}

\newcommand{\lo}[1]{{#1}}
\newcommand{\sh}[1]{}
\newif\ifADDpagenumber
\ADDpagenumbertrue

\allowdisplaybreaks[4]
\usepackage{amsthm}
\newtheorem{theorem}{Theorem}
\newtheorem{lemma}{Lemma}
\newtheorem{corollary}{Corollary}
\newtheorem{remark}{Remark}

\begin{document}

\title{A New Interference-Alignment Scheme for Wireless MapReduce} 

\sloppy
\allowdisplaybreaks[4]

\author{
    \IEEEauthorblockN{Yue Bi\IEEEauthorrefmark{1}\IEEEauthorrefmark{2}, Mich\`ele Wigger\IEEEauthorrefmark{2}, Yue Wu\IEEEauthorrefmark{1}}
    \IEEEauthorblockA{\IEEEauthorrefmark{1}\textit{School of Electronic Information and Electrical Engineering, Shanghai Jiao Tong University}, China; wuyue@sjtu.edu.cn}
    \IEEEauthorblockA{\IEEEauthorrefmark{2} \textit{LTCI, Telecom Paris, IP Paris}, 91120 Palaiseau, France; 
    \{bi, michele.wigger\}@telecom-paris.fr}
}

\maketitle

\ifADDpagenumber
\thispagestyle{plain} 
\pagestyle{plain}
\fi

\begin{abstract}
We consider a full-duplex wireless Distributed Computing (DC) system under the MapReduce framework. New upper  and lower bounds on the optimal tradeoff between Normalized Delivery Time (NDT) and computation load are presented. The upper bound strictly improves over the previous reported upper bounds and is   based on a novel interference alignment  (IA) scheme tailored to the interference cancellation capabilities of MapReduce nodes. 
The lower bound is proved  through   information-theoretic converse arguments. 
\end{abstract}

\section{Introduction}
\emph{Distributed Computing (DC) systems} are computer networks that through task-parallelization reduce execution times  of complex computing tasks such as data mining or computer vision.  MapReduce is a popular such framework and runs in three phases \cite{ng_comprehensive_2021, dean2008mapreduce}. In the first \emph{map phase}, nodes calculate intermediate values (IVA) from  their associated input files. In the following  \emph{shuffle phase}, nodes exchange these IVAs in a way that each node obtains all IVAs required to compute its assigned output function in the  final \emph{reduce phase}. MapReduce is primarily applied to wired systems where it has been noticed that  a significant part of the  MapReduce execution time stems from the IVA \emph{delivery time} during the shuffle phase \cite{dean2008mapreduce, chowdhury2011managing}, and can be reduced through smart coding  \cite{li_fundamental_2018, yan2019storage, xu_new_2021,  chowdhury2011managing, li_unified_2016, zhang_improved_2019}.

MapReduce  systems are becoming important building blocks also in wireless scenarios, such as vehicular networks \cite{vehicular} or distributed e-health applications  \cite{ehealth}, thus creating a need for good wireless  MapReduce coding schemes. 
Similarly to the  wired case  \cite{li_fundamental_2018, yan2019storage, xu_new_2021}, delivery time in wireless MapReduce systems can be decreased by sending appropriate linear combinations of the IVAs, from which  the receiving nodes can extract their desired IVAs by bootstrapping the IVAs that they can compute from their locally stored input files. Further improvements can however be achieved by exploiting specific wireless communication techniques. 

The focus of this paper is on 
the high Signal-to-Noise Ratio (SNR) regime, and on the following two key metrics: 
\begin{itemize}
	\item \emph{Computation load} $\sf r$: This describes the average number of nodes to which each file is assigned. \lo{In other words, it is the ratio of the total number of assigned input files (including replications) normalized by
	the total number of files.}
	\item \emph{Normalized Delivery Time (NDT)} $\Delta$: This is the wireless shuffle duration  normalized by the number of reduce functions and input files and by  the transmission time of a single IVA over a point-to-point channel.
\end{itemize}
We are interested in the minimal NDT for given computation load $\r$, which we call the \emph{NDT-computation tradeoff}.

The NDT of full-duplex interference networks was considered in \cite{li_wireless_2019,bi_dof_2022}, see \cite{yuan_coded_2022} for the half-duplex network. In \cite{li_wireless_2019}, the authors proposed to divide the nodes into groups and let each group  store a subset of the files and apply  one-shot beamforming and  zero-forcing during the shuffle communication. As   shown in \cite{li_wireless_2019}, their scheme is optimal among this class of communication strategies.  In our previous  work  \cite{bi_dof_2022}, we  reduced this NDT  by introducing the IA technique to the shuffle communication in \cite{li_wireless_2019}. 

In this paper, we obtain yet a further NDT reduction by considering the map procedure in \cite{li_fundamental_2018}, where \emph{each set of $\r$ nodes} stores a subset of the files, and  by proposing  a novel 
 IA scheme that is tailored to this file assignment and the interference cancellation capabilities of MapReduce nodes so as to obtain an improved performance compared to standard IA schemes. Our scheme is related to the IA scheme in \cite{hachem_degrees_2018}, which considers a similar file assignment, and to our previous work \cite{bi_dof_2022}, with which it coincides when each node can only store a single file. 

The upper bound on the NDT implied by our new IA-scheme improves over the previously proposed bounds in \cite{li_wireless_2019,bi_dof_2022} whenever  the computation load $\r$ satisfies $1<\r < \left\lceil \frac{\K-1}{2} \right\rceil$. 
Our results thus show that in this regime, beamforming and zero-forcing cannot achieve minimum NDT. 
On the contrary, through an information-theoretic lower bound  on the minimum NDT, in this manuscript we show that for $\r \geq \frac{\K}{2}$ the 
zero-forcing and interference cancellation scheme in \cite{li_wireless_2019} is optimal also without any restriction on the utilized coding scheme.

\textit{Notations:} We use standard notation, and also
 define $[n] \triangleq \{1,2,\ldots,  n\}$ and  $[\mathcal{A}]^n$ as the collection of all the subsets of $\mathcal{A}$ with cardinality $n$. 
%

\section{Wireless MapReduce Framework} \label{sec:WDC_system}

Consider a  distributed computing (DC) system with a fixed number of $\sf K$ nodes labelled $1,\ldots, \sf K$;  a  large number $\sf N$ of input files  $W_1,\ldots, W_{\sf{N}}$; and $\sf{K}$ output functions $\phi_1, \ldots, \phi_{\sf K}$ mapping the input files to the desired computations. 

A \emph{MapReduce} System decomposes the output functions as: 
\begin{equation}
\phi_q( W_1,\ldots, W_{\sf N})= v_q( a_{q,1}, \ldots, a_{q,\sf N}), \qquad q \in [\sf{K}],
\end{equation}
where $v_q$ is called \emph{reduce function} and $a_{q,p}$ the\emph{intermediate value (IVA)} calculated from  file $W_p$ through a \emph{map function}
\begin{equation}
a_{q,p}= u_{q,p}(W_p), \qquad p\in[\sf N].
\end{equation} 
 IVAs are  independent  with $\sf{A}$ i.i.d. bits. 

The MapReduce framework has 3 phases:

\textbf{Map phase}: A subset of all input files $\mathcal{M}_k \subseteq [\sf N]$ is assigned to each  node $k \in [\sf K]$. Node $k$ computes all IVAs $\{a_{q,p} \colon p \in \mathcal{M}_k, \, q \in [\sf K]\}$ associated with these input files. Notice that the set $\{\mathcal{M}_k\}_{k \in [\sf K]}$ is a design factor.
   
 \textbf{Shuffle phase:} Computation of the $k$-th output function is assigned to the $k$-th node. 
 
    The $\sf K$ nodes in the system communicate over $\sf T$ uses of a wireless network in a full-duplex mode to exchange the . 
     missing  IVAs for the computations of their assigned output functions.  So, node $k\in[\sf  K]$  produces  complex channel inputs of the form
    \begin{equation}\label{eq:comp_encoding}
    \vect{X}_k  \triangleq (X_k(1),\ldots, X_k(\sf T))^{\intercal}= f_k^{(\sf  T)}\left(\{a_{1, p}, \ldots,  a_{\sf K, p}\}_{p \in \mathcal M_k}\right),
    \end{equation}  
by means of an encoding function $f_k^{(\sf T)}$ on appropriate domains and so that the inputs satisfy the  block-power constraint
\begin{IEEEeqnarray}{rCl}\label{eq:power}
	\frac{1}{\sf T} \sum_{t=1}^{\sf T}     \mathbb{E}\left[|X_{k}(t)|^2\right] \leq\sf  P, \qquad k\in[\sf K].
\end{IEEEeqnarray}
    Given the full-duplex nature of the network, Node $k$  also observes  the complex channel outputs
    \begin{equation}\label{eq:channel}
    Y_{k}(t)= \sum_{k' \in [\sf K]\backslash \{k\}}H_{k,k'} (t) X_{k'}(t) + Z_{k}(t), \quad t\in[\sf T],
    \end{equation}
    where the sequences of complex-valued channel coefficients $\{H_{k,k'}(t)\}$  and standard circularly symmetric Gaussian noises $\{Z_{k}(t)\}$ are both i.i.d. and  independent of each other and of all other channel coefficients and noises.
    
Based on its outputs $\vect{Y}_{k} \triangleq (Y_k(1),\ldots, Y_k(\sf T))^{\intercal}$ and the IVAs $\{a_{q,p} \colon p \in \mathcal{M}_k, \; q \in [\sf K]\}$  it computed during the Map phase, Node~$k$  decodes the missing IVAs $\{a_{k,p}\colon p \notin \mathcal{M}_k\}$ required to  compute its assigned output functions $\phi_k$ as:
\begin{equation}\label{eq:comp_decoding}
\hat{a}_{k,p} = g_{k,p}^{(\sf  T)}\left(\{a_{1, i}, \ldots, a_{\sf K, i}\}_{i \in \mathcal M_k}, \vect{Y}_{k}\right),  \quad p \notin \mathcal{M}_k. 
\end{equation}
    
 \textbf{Reduce phase:} Each node $k\in[\K]$ applies  reduce functions $\phi_k(\cdot)$ to the appropriate  IVAs calculated during the Map phase  or decoded in the Shuffle phase.

The performance of the distributed computing system is  measured  in terms of its \emph{computation load}
\begin{equation}\label{eq:r}
   \sf r \triangleq \sum_{k \in [\sf K]} \frac{|\mathcal{M}_k|}{\sf N},
\end{equation}
and the \emph{normalized delivery time (NDT)}
\begin{equation}\label{eq:Delta}
    \sf \Delta \triangleq \varliminf_{\sf P \rightarrow \infty} \varliminf_{\sf A \to \infty} \frac{\sf T}{\sf A \cdot  \sf K\cdot \sf N} \cdot \log \sf P.
\end{equation}

We focus on the \emph{fundamental NDT-computation tradeoff} $\Delta^*(\sf r)$, which is defined as  the  infimum over all values of  $\Delta$ satisfying \eqref{eq:Delta} for some choice of file assignments $\{\mathcal{M}_k\}$  and sequence (in $\sf T$) of encoding and decoding functions $\{f_k^{(\sf T)}\}$ and $\{g_{k,p}^{(\sf T)}\}$ in \eqref{eq:comp_encoding} and \eqref{eq:comp_decoding}, all depending on $\sf A$ so that the probability of  IVA decoding error
\begin{equation}\label{eq:error_computing}
 \textnormal{Pr}\bigg[ \bigcup_{ \substack{k\in[{\sf K}]}}\; \bigcup_{p \notin\mathcal{M}_k}  \hat{a}_{k,p}\neq a_{k,p} \bigg] \to 0 \quad \textnormal{as} \quad \sf A \to \infty.
\end{equation} 


\subsection{Sufficiency of Symmetric File Assignments }\label{sec:suff}

Our model exhibits a perfect symmetry between the various nodes in the network because the channels from any  Tx-node to any Rx-node are independent and have identical statistics. 
The optimal NDT-computation tradeoff can therefore be achieved by a symmetric file assignment where any subset of nodes $\mathcal{T}\subseteq [\K]$ of size $i$ is assigned the same number of  files to be stored at all nodes in $\mathcal{T}$. \lo{In fact, any non-symmetric file assignment can be symmetrized without decreasing the NDT-computation tradeoff. It suffices to time-share $\K!$ instances of the original scheme for a IVA size that is also multiplied by $\K!$, where in each instance the   $\K$ nodes are relabeled according  to a different permutation. The resulting scheme has a symmetric file assignment and achieves the same NDT-computation tradeoff as the original scheme because both $\mathsf{T}$ and $\mathsf{A}$ are multiplied by $\K!$ while the other parameters remain unchanged and because the new scheme still satisfies \eqref{eq:error_computing}.}
\sh{In fact, any non-symmetric file assignment can be symmetrized by time-sharing.}


\subsection{Relation to the Sum-DoF with $\r$-fold Cooperation}\label{sec:relation}
A well-studied property of wireless networks is the \emph{Sum Degrees of Freedom (sum-DoF)}, which characterizes the maximum throughput of a network. In this work we are specifically interested in the sum-DoF that one can achieve over the wireless network described by \eqref{eq:channel}, when the inputs are subject to the average power constraints \eqref{eq:power} and any set of $\r$ nodes $\mathcal{T}\in [\K]^\r$ has a message $M_{\mathcal{T}}^j$ that it  wishes to convey to Node $j$, for any $j\in [\K]\backslash \mathcal{T}$. Each message $M_{\mathcal{T}}^j$ is uniformly distributed over a set $[2^{n R_{\mathcal{T}}^j}]$ and a rate-tuple $( R_{\mathcal{T}}^j \colon \mathcal{T}\in [\K]^\r, \; j\in [\K]\backslash \mathcal{T})$ is called achievable if there exists a sequence of encoding and decoding functions such that the probabilities of error tend to 0 in the asymptotic regime of infinite blocklengths. The sum-DoF is then defined as 
\begin{equation} 
 \SDoF(\r) \triangleq  \sup \varlimsup_{P\to \infty} \frac{ \sum_{  \mathcal{T}\in [\K]^\r, \; j\in [\K]\backslash \mathcal{T}} R_{\mathcal{T}}^j(\sf P)  }{ \frac{1}{2}\log \sf P},
\end{equation}
where the supremum is over  sequences $\{( R_{\mathcal{T}}^j(\sf P) \colon \mathcal{T}\in [\K]^\r, \; j\in [\K]\backslash \mathcal{T})\}_{\sf P >0}$  so that for each $\sf P>0$ each tuple $(R_{\mathcal{T}}^j(\sf P) \colon \mathcal{T}\in [\K]^\r, \; j\in [\K]\backslash \mathcal{T})$ is achievable under power $\sf P$. 

\begin{lemma} \label{lem:E}
For any $\r \in [\K]$: 
\begin{equation}\label{eq:DSDof}
    \sf \Delta(\r) \geq  \left(1-\frac{\r}{\K}\right) \frac{1}{ \SDoF(\r)}.
\end{equation}
\end{lemma} 
\begin{IEEEproof}\sh{See the long version \cite{bi2022bounds},}\lo{See Appendix~\ref{app:lem1},} but the idea is well known and also used in \cite{li_wireless_2019} and \cite{bi_dof_2022}.\end{IEEEproof}

\section{New IA Scheme} \label{sec:scheme}
In view of Lemma~\ref{lem:E},  in this section we present a scheme achieving a high $\SDoF(\r)$ over the  wireless network. 
We start with an example.
\subsection{Example 2:  $\K=4, \r=2$}\label{sec:examples}

Consider $\r=2$ and  $\K=4$ nodes. 
In this case our scheme transmits  18 different messages depicted in \eqref{eq:messages}. Here, Message $M_{k,\mathcal{T}}^j$ is a message that is known by the set of nodes $\mathcal{T}$ and intended to Node $j\notin \mathcal{T}$. (Since we consider $\r$-fold cooperation, we have $|\mathcal{T}|=2$.) 

 Message $M_{k,\mathcal{T}}^j$ is only transmitted by a single Node $k \in \mathcal{T}$. The remaining nodes in $\mathcal{T}\backslash \{k\}$ only  
 exploit their knowledge of   $M_{k,\mathcal{T}}^j$ to  cancel the   transmission from their receive signal. 

 Notice that for certain sets $\mathcal{T}$ and receive nodes $j\notin \mathcal{T}$ our scheme sends two messages to the same node $j$:  $M_{k_1,\mathcal{T}}^j$ and $M_{k_2,\mathcal{T}}^j$  for $k_1 \neq k_2$. \lo{(In  \eqref{eq:messages} the two messages $ M_{1, \{1,4\}}^{2}$ and $M_{4, \{1,4\}}^{2}$ for example have this form.)} These messages   $M_{k_1,\mathcal{T}}^j$ and $M_{k_2,\mathcal{T}}^j$  actually represent two independent submessages of Message $M_{\mathcal{T}}^j$ as we defined it in Section~\ref{sec:relation}. For the sets $\mathcal{T}$ and Nodes $j\notin \mathcal{T}$  for which there exists only a single Message $M_{k,\mathcal{T}}^j$, this message is really the message $M_{\mathcal{T}}^j$ as we defined it in Section~\ref{sec:relation}.  Since our interest is on the Sum-DoF,   distinction between submessages and messages  is not relevant.

In our scheme, to Node~1 we send messages
\begin{subequations}\label{eq:messages}
\begin{IEEEeqnarray}{rCl}
M_{2, \{2,3\}}^{1}, \ M_{3, \{2,3\}}^{1},\ M_{2, \{2,4\}}^{1}, \  M_{3, \{3,4\}}^{1} ;\IEEEeqnarraynumspace
\end{IEEEeqnarray}
to Node 2 we send messages
\begin{IEEEeqnarray}{rCl}
M_{1, \{1,3\}}^{2}, \ M_{3, \{1,3\}}^{2},  \ M_{1, \{1,4\}}^{2}, \ M_{4, \{1,4\}}^{2} \nonumber \\
   M_{3, \{3,4\}}^{2} \ M_{4, \{3,4\}}^{2}  ; \IEEEeqnarraynumspace
\end{IEEEeqnarray}
to Node 3 we send messages
\begin{IEEEeqnarray}{rCl}
M_{1, \{1,2\}}^{3}, \ M_{2, \{1,2\}}^{3}, \ M_{1, \{1,4\}}^{3}, \ M_{4, \{1,4\}}^{3}, \nonumber \\
 M_{2, \{2,4\}}^{3}, \ M_{4, \{2,4\}}^{3} ;\IEEEeqnarraynumspace
\end{IEEEeqnarray}
and to Node 4 we send messages
\begin{IEEEeqnarray}{rCl}
M_{1, \{1,2\}}^{4}, \ M_{2, \{1,2\}}^{4}, \ M_{1, \{1,3\}}^{4}, \ M_{3, \{1,3\}}^{4}, \nonumber \\ M_{2, \{2,3\}}^{4} , \ M_{3, \{2,3\}}^{4}.\IEEEeqnarraynumspace
\end{IEEEeqnarray}
\end{subequations}
Node 
$\K=4$  does not send any message to the first Node $1$. \lo{(This omission allows to reuse some of the precoding matrices and achieve an improved Sum-DoF.)}
We encode each message $M_{k,\mathcal{T}}^j$  into a Gaussian codeword  $\vect b_{k,\mathcal{T}}^j$ and  use  IA  with three  precoding matrices $\mat{U}_{\{2,3\}}, \mat{U}_{\{2,4\}}$, and $\mat{U}_{\{3,4\}}$. Matrix $\mat{U}_{\{2,3\}}$ is used to send   codewords
\begin{IEEEeqnarray}{rCl}
 \ \vect{b}_{1, \{1,3\}}^{2},  \ \vect{b}_{1, \{1,2\}}^{3}, \  \vect{b}_{4,\{3,4\}}^2, \  \vect{b}_{4, \{2,4\}}^{3}, \label{eq:I01}  \\
\vect{b}_{2, \{2,3\}}^{1},  \ \vect{b}_{3, \{2,3\}}^{1}, \ \vect{b}_{3,\{1,3\}}^2, \ \vect{b}_{2,\{1,2\}}^3, \label{eq:l1}
\end{IEEEeqnarray}
matrix $\mat{U}_{\{2,4\}}$ for codewords 
\begin{IEEEeqnarray}{rCl}
 \ \vect{b}_{1, \{1,4\}}^{2},  \ \vect{b}_{1, \{1,2\}}^{4}, \  \vect{b}_{3,\{3,4\}}^2, \  \vect{b}_{3, \{2,3\}}^{4},   \label{eq:I02}\\
\vect{b}_{2, \{2,4\}}^{1},  \ \vect{b}_{2,\{1,2\}}^4, \ \vect{b}_{4,\{1,4\}}^2, \label{eq:l2}
\end{IEEEeqnarray}
and  matrix $\mat{U}_{\{3,4\}}$ for codewords 
\begin{IEEEeqnarray}{rCl}
 \ \vect{b}_{1, \{1,4\}}^{3},  \ \vect{b}_{1, \{1,3\}}^{4}, \  \vect{b}_{2,\{2,4\}}^3, \  \vect{b}_{2, \{2,3\}}^{4},  \label{eq:I03} \\
\vect{b}_{3, \{3,4\}}^{1}, \ \vect{b}_{4,\{1,4\}}^3, \ \vect{b}_{3,\{1,3\}}^4.\label{eq:l3}
\end{IEEEeqnarray}

\begin{remark}The  choice of precoding matrices is inspired by \cite{hachem_degrees_2018} where  Message $M_{k, \mathcal{T}}^{j}$ is precoded by the matrix $\mat{U}_{\mathcal{R}}$ for $\mathcal{R}=\mathcal{T} \backslash \{j\} \cup \{k\}$. \sh{Since any node in $\mathcal{R}$ is either interested in learning Message $M_{k, \mathcal{T}}^{j}$ or it can compute it itself and remove the interference, any node only experiences interference from precoding matrices $\mat{U}_{\mathcal{R}}$ for which $j\notin \mathcal{R}$.}  \lo{The idea behind the choice of precoding matrices in \cite{hachem_degrees_2018} is that  any node in $\mathcal{R}$ is either interested in learning Message $M_{k, \mathcal{T}}^{j}$ or it can compute it itself and remove the interference from its receive signal. A given node $j$ thus only experiences interference from precoding matrices $\mat{U}_{\mathcal{R}}$ for which $j\notin \mathcal{R}$.} In our IA scheme, we omit precoding matrices $\mat{U}_{\mathcal{R}'}$ for sets $\mathcal{R}'$ containing index $1$, 
and instead use the precoding matrix $\mat{U}_{\mathcal{R}}$ also to send  \begin{equation}\label{eq:form}
\vect b_{k,\mathcal{R}}^1, \qquad \vect b_{k,\mathcal{R}\cup \{1\} \backslash \{j\}}^j, \qquad \forall j,k\in \mathcal{R}, \; j\neq k,
\end{equation}
see the codewords indicated in  \eqref{eq:l1}, \eqref{eq:l2}, \eqref{eq:l3}. 
\end{remark}
We illustrate our assignment of the precoding matrices also in Table~\ref{table:K3r2}. The entries in column 1 or in rows $\{1,2\}, \{1,3\}, \{1,4\}$ correspond to two submessages $M_{k_1,\mathcal{T}}^j$ and $M_{k_2,\mathcal{T}}^j$, where $k_1$ and $k_2$ denote the two entries in $\mathcal{T}$. For all other entries  in Table~\ref{table:K3r2} not equal to ``x", we have only one message per precoding matrix, see \eqref{eq:I01}, \eqref{eq:I02}, and \eqref{eq:I03}. 

\begin{table}

\caption{Messages $M_{k,\mathcal{T}}^j$ precoded by the three precoding  matrices  $\mat{U}_{\{2,3\}}$, $\mat{U}_{\{2,4\}}$, and $\mat{U}_{\{3,4\}}$. }
\label{table:K3r2}
\centering
\begin{tabular}{c  ||  c  | c | c| c  }
$\mathcal{T} \; \backslash \; j$ &1 & 2 & 3 &4 \\ [0.5ex] 
\hline\hline 
$\{1,2\}$& x &  x& $\mat{U}_{\{2,3\}}$ &$\mat{U}_{\{2,4\}}$ \\[1.1ex]
$\{1,3\}$& x & $\mat{U}_{\{2,3\}}$ & x &$\mat{U}_{\{3,4\}}$ \\[1.1ex]
$\{1,4\}$& x &$\mat{U}_{\{2,4\}}$ &$\mat{U}_{\{3,4\}}$ & x\\[1.1ex]
$\{2,3\}$& $ \mat{U}_{\{2,3\}}$&x &  x&$\mat{U}_{\{2,4\}}$,  \\
&&&&$\mat{U}_{\{3,4\}}$\\[1.1ex]
$\{2,4\}$& $\mat{U}_{\{2,4\}}$ &  x& $ \mat{U}_{\{2,3\}}$,  &x\\
&&&$\mat{U}_{\{3,4\}}$ &\\[1.1ex]
$\{3,4\}$&$\mat{U}_{\{3,4\}}$ &$ \mat{U}_{\{2,3\}}$, $\mat{U}_{\{2,4\}}$  &x&x\\
\end{tabular}
\vspace{-2mm}

\end{table}

During the shuffling phase, \lo{Nodes~1--4  send the following signals. }
Node 1 sends:  
\begin{IEEEeqnarray}{rCl}
\vect{X}_1& =& \mat{U}_{\{2,3\}}  \left(\vect{b}_{1, \{1,3\}}^{2}+   \vect{b}_{1, \{1,2\}}^{3}\right)  \nonumber \\
&& + \mat{U}_{\{2,4\}}  \left(  \vect{b}_{1, \{1,4\}}^{2}+ \vect{b}_{1, \{1,2\}}^{4} \right)  \nonumber \\
&&+ \mat{U}_{\{3,4\}} \left( \vect{b}_{1, \{1,4\}}^{3} + \vect{b}_{1, \{1,3\}}^{4}\right). 
\end{IEEEeqnarray}
\lo{Node 2 sends: 
\begin{IEEEeqnarray}{rCl}
\vect{X}_2 & =& \mat{U}_{\{2,3\}}  \left(\vect{b}_{2, \{2,3\}}^{1}+ \vect{b}_{2,\{1,2\}}^3\right)  \nonumber \\
&& + \mat{U}_{\{2,4\}}  \left( \vect{b}_{2, \{2,4\}}^{1}+ \vect{b}_{2,\{1,2\}}^4 \right)  \nonumber \\
&&+ \mat{U}_{\{3,4\}} \left(   \vect{b}_{2,\{2,4\}}^3+ \vect{b}_{2, \{2,3\}}^{4}\right). 
\end{IEEEeqnarray}
Node 3 sends:
\begin{IEEEeqnarray}{rCl}
\vect{X}_3 & =& \mat{U}_{\{2,3\}}  \left(\vect{b}_{3, \{2,3\}}^{1}+\vect{b}_{3,\{1,3\}}^2\right)  \nonumber \\
&& + \mat{U}_{\{2,4\}}  \left( \vect{b}_{3,\{3,4\}}^2 + \vect{b}_{3, \{2,3\}}^{4} \right)  \nonumber \\
&&+ \mat{U}_{\{3,4\}} \left( \vect{b}_{3, \{3,4\}}^{1}+ \vect{b}_{3,\{1,3\}}^4\right). 
\end{IEEEeqnarray} }
\sh{Nodes 2 and 3 send similar signals based on the assigned codewords.}
Node 4 sends: 
\begin{IEEEeqnarray}{rCl}
\vect{X}_4 & =& \mat{U}_{\{2,3\}}  \left(\vect{b}_{4,\{3,4\}}^2+  \vect{b}_{4, \{2,4\}}^{3}\right)  \nonumber \\
&& + \mat{U}_{\{2,4\}} \vect{b}_{4,\{1,4\}}^2+ \mat{U}_{\{3,4\}}    \vect{b}_{4,\{1,4\}}^3 . 
\end{IEEEeqnarray}

Eeach  node  subtracts all the interference of the signals that it can compute itself. 
For example, Node~2 thus  constructs: 
\begin{IEEEeqnarray}{rCl}
\vect{Y}_2'& =& \underbrace{\mat{H}_{2,1} \mat{U}_{\{2,3\}}  \vect{b}_{1, \{1,3\}}^{2}+ \mat{H}_{2,1} \mat{U}_{\{2,4\}}  \vect{b}_{1, \{1,4\}}^{2}  }_{\textnormal{desired signal}} \nonumber \\
& & \underbrace{+  \mat{H}_{2,3}  \mat{U}_{\{2,3\}}  \vect{b}_{3, \{1,3\}}^{2}  + \mat{H}_{2,3}\mat{U}_{\{2,4\}}  \vect{b}_{3, \{3,4\}}^{2} }_{\textnormal{desired signal}}\nonumber\\
& & \underbrace{+  \mat{H}_{2,4}  \mat{U}_{\{2,3\}}  \vect{b}_{4, \{3,4\}}^{2}  + \mat{H}_{2,4}\mat{U}_{\{2,4\}}  \vect{b}_{4, \{1,4\}}^{2} }_{\textnormal{desired signal}}\nonumber\\
& &+\mat{H}_{2,1} \mat{U}_{\{3,4\}} \left(   \vect{b}_{1,\{1,4\}}^3+ \vect{b}_{1, \{1,3\}}^{4}\right)\nonumber \\
&&  +\mat{H}_{2,3} \mat{U}_{\{3,4\}}  \left( \vect{b}_{3,\{3,4\}}^1 + \vect{b}_{3, \{1,3\}}^{4} \right) \nonumber \\
& &  +\mat{H}_{2,4} \mat{U}_{\{3,4\}}  \vect{b}_{4,\{2,4\}}^3  +\vect{Z}_2.
\end{IEEEeqnarray}

 Node 2's desired signals are all precoded by precoding matrices $\mat{U}_{\{2,3\}}$ and $\mat{U}_{\{2,4\}}$, while all interference signals are precoded by matrix $\mat{U}_{\{3,4\}}$. Similar observations hold for Nodes 3 and 4. Desired and interference signals at Node 1 instead are precoded by all three precoding matrices. 

%
%

The IA matrices $\mat{U}_{\{2,3\}}$, $\mat{U}_{\{2,4\}}$, and $\mat{U}_{\{3,4\}}$ are constructed based on the IA idea in \cite{jafar_degrees_2008} taking into account the channel matrices that premultiply the IA matrices in the interference signals of $\vect{Y}_1', \vect{Y}_2', \vect{Y}_3', \vect{Y}_4'$. Specifically: 
\begin{IEEEeqnarray}{rCl}\label{eq:U_cR}
\mat{U}_{\cR}\triangleq  \left[ \prod_{\mat{H} \in {\mathcal{H}}_{\cR} }
\mat{H}^{{\alpha_{\cR,\mat{H}}}}\cdot  \boldsymbol{\Xi}_{\cR}\colon  \,\; 
\forall \boldsymbol{\alpha}_{\cR} \in  [\eta]^{4} \right], \quad 
\end{IEEEeqnarray}
where  each column of the matrix is constructed using  a different exponent-vector $\boldsymbol{\alpha}_{\mathcal{R}}=( \alpha_{\mathcal{R},\mat{H}} \colon \mat{H} \in \mathcal{H}_{\mathcal{R}})\in  [\eta]^{4}$;
$\eta$ is a large number depending on the blocklength $\sf T$ that tends to $\infty$ with $\sf T$;  $\boldsymbol{\Xi}_{\cR}$ are i.i.d.  random vectors drawn according to a continuous distribution, and
\begin{IEEEeqnarray}{rCl}
	\label{eq:set_G}
	\mathcal{H}_{\{2,3\}} & \triangleq& \big\{ \mat{H}_{1,4} , \ \mat{H}_{4,1} , \ \mat{H}_{4,2},   \ \mat{H}_{4,3} \big\},\\
		\mathcal{H}_{\{2,4\}} & \triangleq& \big\{ \mat{H}_{1,3} , \ \mat{H}_{3,1} , \ \mat{H}_{3,2},   \ \mat{H}_{3,4} \big\},\\
	\mathcal{H}_{\{3,4\}} & \triangleq& \big\{ \mat{H}_{1,2} , \ \mat{H}_{2,1} , \ \mat{H}_{2,3},   \ \mat{H}_{2,4} \big\}.
\end{IEEEeqnarray}

By this choice of the precoding matrices, all interference signals at a Node $2$  will lie 
in the columnspace of the matrix 
\begin{IEEEeqnarray}{rCl}\label{eq:W_cR}
\mat{W}_{\{3,4\}}\triangleq  \left[ \prod_{\mat{H} \in {\mathcal{H}}_{\{3,4\}} }
\mat{H}^{{\alpha_{\cR,\mat{H}}}}\cdot  \boldsymbol{\Xi}_{\{3,4\}}\colon  \,\; 
\forall \boldsymbol{\alpha}_{\cR} \in  [\eta+1]^{4} \right], \nonumber
\end{IEEEeqnarray}
while the desired signals will be separable from each other and from this interference space. 

Since $\vect{Y}_2'$ has 6 signal spaces for the $6$ desired signals and only a single interference space, our scheme achieves DoF $6/7$ to Node~$2$. 
  Similar considerations hold for Nodes $3$ and $4$. Node~$1$ has 3  interference spaces (one for each IA precoding matrix) and only 4 signal spaces, and we thus achieve DoF  $4/7$ to Node 1. 
  The SumDoF achieved by the scheme is thus: 
 $\SDoF = 22/7$.
 
For comparison, notice that the IA scheme in \cite{hachem_degrees_2018} uses the additional precoding matrices $\mat{U}_{\{1,2\}}$, $\mat{U}_{\{1,3\}}$, and $\mat{U}_{\{1,4\}}$. The interference space at Node 2 then consists of $\mat{W}_{\{3,4\}}$  and additionally of the similar matrices $\mat{W}_{\{1,2\}}$ and  $\mat{W}_{\{1,4\}}$. Thus, only DoF $6/(6+3)=2/3$ is  achievable to Node 2, as well as to all other nodes, leading to the reduced Sum-DoF $12/3$.
 \subsection{The General Scheme} \label{sec:ach}

\renewcommand{\mu}{\sf T}
Fix  $\eta \in \mathbb{Z}^+$ (which we shall let  tend to $\infty$) and let 
\begin{IEEEeqnarray}{rCl}\Gamma & \triangleq & \sf K  \cdot (\sf K- \sf r -1)\\
	\sf T& \triangleq &(\sf K -2) 
	\cdot \binom{\sf K -2}{\sf r-1 }\cdot \eta^\Gamma + \binom{\sf K-1}{\sf r} \cdot (\eta+1)^\Gamma. \label{eq:T}\IEEEeqnarraynumspace
\end{IEEEeqnarray}

We send the following messages to any Node $j\in [\K]\backslash \{1\}$:
\begin{IEEEeqnarray}{rCl}
\left\{ M_{k, \mathcal{T}}^j \colon \mathcal{T}\in [ [\K]\backslash \{j\} ]^{\r}, \ k \in \mathcal{T}\right \}
\end{IEEEeqnarray}
and to Node 1 we send messages
\begin{IEEEeqnarray}{rCl}
\left\{M_{k, \mathcal{T}}^1 \colon \mathcal{T}\in [ [\K]\backslash \{1\}]^{\r}, \ k \in \mathcal{T}\backslash \{\K\}\right \}.
\end{IEEEeqnarray}
Thus, as in the examples of the previous section, the last node $\K$ does not send any message to the Node $1$.

For each message,  construct a Gaussian codebook of power $\sf P /{\sf K-1 \choose  \sf r}$ and length $\eta^{\Gamma}$ to encode each Message ${M}_{k,\cT}^{j}$ into a codeword $\vect b_{k,\cT}^j$.
As in the previous sections, we shall use a linear precoding scheme,  
and thus Node $i \in [\sf K]$ can  mitigate the interference caused by the codewords  
\begin{equation}
 \big\{\vect{b}_{k,\cT}^{j} \big\}_{\forall \cT \colon i \in \cT}. 
\end{equation}
Thus, for each set $\cR\in[[\K]]^{\r}$, without causing non-desired interference to nodes in $\cR$, we can  use the same precoding matrix $\mat{U}_{\cR}$   for all the codewords:
\begin{equation}\label{eq:cod1}
 \big\{\vect{b}_{k,\cR  \cup  \{k\} \backslash \{j\}}^{j} \big\}_{\substack{k \in[\sf K]\backslash \cR \\ j\in \cR}} .
 \end{equation} 
This idea was already used in the related works \cite{hachem_degrees_2018,yuan_coded_2022}. In contrast to these previous works, here we do not introduce the precoding matrices $\mat{U}_{\cR}$ for sets $\cR$ containing $1$ and instead 
 we use   matrix $\mat{U}_{\cR}$, for $1\notin \mathcal{R}$,  also to precode the codewords 
\begin{equation}\label{eq:MRk}
\left\{\vect{b}_{k,\cR  \cup  \{1\} \backslash \{j\}}^{j} \right\}_{j,k\in\mathcal{R}, j\neq k}  \quad \cup \quad \left\{\vect{b}_{1,\cR  \cup  \{1\} \backslash \{j\}}^{j} \right\}_{j\in\mathcal{R}},
\end{equation}
and
\begin{equation}\label{eq:cod3}
 \{ \vect{b}_{k,\cR}^{1} \}_{k\in\mathcal{R}\backslash \{\K\}}.
\end{equation}
All non-intended nodes in $\cR$ can subtract these interferences from their receive signals because they know the codewords. This trick 
allows us to reduce the dimension of the interference space and thus improve performance.

Table~\ref{table:K6r2_one} illustrates  which codesymbols $\vect{b}_{k,\mathcal{T}}^{j}$  are premultiplied by the precoding matrix  $\mat{U}_{\{2,3,4\}}$,  when $\r=3$ and $\K \geq 5$. The entries in rows $\mathcal{T}$ containing index $1$ correspond to $\r=3$ different codewords $\vect{b}_{k,\mathcal{T}}^j$, one for each $k\in\mathcal{T}$, see \eqref{eq:MRk}. Similarly, the entry in column-$1$ and row $\{2,3,4\}$  corresponds to the $\r$ codewords $\vect{b}_{k,\{2,3,4\}}^1$, for each $k\in\{2,3,4\}$. Any other entry of the table showing $\mat{U}_{\{2,3,4\}}$ corresponds to a single codeword $\vect{b}_{k,\mathcal{T}}^j$, where $k$ is the single element in $\mathcal{T}\backslash\{2,3,4\}$.  Similar tables can be drawn for all pairs $(k_1,k_2)\in [\K]$, where recall  that  node $\K$ does not send any information to Node 1.

\begin{table}
\caption{The codewords $\vect{b}_{k,\mathcal{T}}^{j}$  premultiplied by  $\mat{U}_{\{2,3,4\}}$.} 
\centering
\begin{tabular}{c  ||  c  | c | c| c |c }
$\mathcal{T}\; \backslash \;  j$ &1 & 2 & 3 &4  \\[0.5ex] 
\hline\hline 
$\{1,2,3\}$& x &  x&x&  $\mat{U}_{\{2,3,4\}}$  \\
$\{1,2,4\}$& x & x& $ \mat{U}_{\{2,3,4\}}$  &  x  \\
$\{1,3,4\}$& x & $ \mat{U}_{\{2,3,4\}}$ & x&x \\
$\{2,3, 4\}$&$ \mat{U}_{\{2,3,4\}}$& x&x &  x\\
$\{2,3,5\}$&  o& x& x&$ \mat{U}_{\{2,3,4\}}$\\
$\vdots $& \vdots & \vdots&$ \vdots$&$ \vdots$ \\
$\{2,3,\K\}$& o &x &x& $ \mat{U}_{\{2,3,4\}}$ \\
$\{2,3,5\}$&  o& x&$ \mat{U}_{\{2,3,4\}}$& x\\
$\vdots $& \vdots & \vdots&$ \vdots$& $ \vdots$\\
$\{2,3,\K\}$& o &x& $ \mat{U}_{\{2,3,4\}}$ & x\\
$\{3,4,5\}$&  o&$ \mat{U}_{\{2,3,4\}}$& x&x\\
$\vdots $& \vdots & \vdots&$ \vdots$&$ \vdots$ \\
$\{3,4,\K\}$& o & $ \mat{U}_{\{2,3,4\}}$& x &x\\
\end{tabular}
\label{table:K6r2_one}
\end{table}


\textit{Encoding:}
Define the $\mu$-length vector of channel  inputs
$\vect{X}_k  \triangleq (X_{k}(1),\ldots,X_{k}(\mu) )^\intercal$ for each Node~$k$ and set: 
\begin{IEEEeqnarray}{rCl}
\vect{X}_{1} &=& \sum_{\cR \in [[\sf K]\backslash \{1\}]^{\sf r}} \; \sum_{j\in \cR}  \mat{U}_{\cR}  \vect{b}_{1,\cR\cup \{1\}\backslash \{j\}}^j , \label{eq:X1}\\
\vect{X}_{k} &=& \sum_{\cR \in [[\sf K]\backslash \{1, k\}]^{\sf r}} \;  \sum_{j\in \cR}  \mat{U}_{\cR} \vect{b}_{k,\cR\cup \{k\}\backslash \{j\}}^j
\nonumber \\
&& + \sum_{ \substack{\cR \in [[\sf K]\backslash \{k\}]^{\sf r} \colon \\ 1 \in \cR}}  \sum_{j\in \cR}   \mat{U}_{\cR \cup \{k\} \backslash \{1\}}  \vect{b}_{k,\cR\cup \{k\}\backslash \{j\}}^j, \nonumber\\[-2ex]
&& \hspace{4.1cm} k\in[\sf K-1]\backslash\{1\},\IEEEeqnarraynumspace  \label{eq:X2}\\
\vect{X}_{\sf K} &=&  \sum_{\cR \in [[\sf K-1]\backslash \{1\}]^{\sf r}} \;  \sum_{j\in \cR}  \mat{U}_{\cR} \vect{b}_{\sf K,\cR\cup \{\sf K\}\backslash \{j\}}^j \nonumber \\
&& + \sum_{ \substack{\cR \in [[\sf K-1]]^{\sf r} \colon \\ 1 \in \cR}}  \sum_{j\in \cR\backslash \{1\}}   \mat{U}_{\cR \cup \{\sf K\} \backslash \{1\}}  \vect{b}_{\sf K,\cR\cup \{\sf K\}\backslash \{j\}}^j , \IEEEeqnarraynumspace
\label{eq:XK}
\end{IEEEeqnarray}
where  we shortly describe  matrices $\{\mat{U}_{\cR}\}_{\cR \in [[\sf K]\backslash\{1\}]^{\sf r}}$.

\textit{Decoding:}
After receiving the respective sequence of $\mu$ channel  outputs $\vect{Y}_j\triangleq (Y_{j,1},\ldots, Y_{j,\mu})$, for $j \in [\sf K]$, each node removes the influence of the codewords corresponding to the messages  that it can compute itself. The  nodes' ``cleaned" signals  can then   be written as: 
\begin{subequations}\label{eq:cleaned}
\begin{IEEEeqnarray}{rCl}
{\vect{Y}}_{1} '&=&\underbrace{ \sum_{\substack{\cR \in [[\sf K]]^{\sf r}\colon \\ 1 \in \cR}} \sum_{\substack{k\in [\sf K-1]\backslash \cR}}\mat{H}_{1, k}\mat{U}_{\cR \cup \{k\}\backslash\{1\} } \vect{b}_{k,\cR\cup \{k\}\backslash \{1\}}^1}_{\text{desired signal}} \nonumber\\
&&+ \sum_{\cR \in [[\sf K]\backslash\{1\}]^{\sf r}} \;\; \sum_{\substack{k \in [\sf K]\backslash \cR}} \mat{H}_{1, k}\mat{U}_{\cR} \vect{v}_{\cR, k} 
 + \vect{Z}_1,\label{eq:Y1}\\
{\vect{Y}}_{j} '&=& \underbrace{\sum_{\substack{\cR \in [[\sf K]\backslash\{1\} ]^{\sf r}\colon \\ j \in \cR }} \; \;\sum_{\substack{k\in[\sf K]\backslash \cR}} \mat{H}_{j,k}\mat{U}_{\cR}\vect{b}_{k,\cR\cup \{k\}\backslash \{j\}}^j}_{\text{desired signal}} \nonumber \\
&&+ \underbrace{\sum_{\substack{\cR \in [[\sf K]]^{\sf r}\colon \\ 1, j \in \cR }}\sum_{\substack{k\in[\sf K]\backslash \cR}}  \mat{H}_{j,k}\mat{U}_{\cR \cup \{k\}\backslash \{1\}}\vect{b}_{k,\cR}^j}_{\text{desired signal}} \nonumber \\
&&+ \sum_{\substack{\cR \in [[\sf K]\backslash\{1\} ]^{\sf r}\colon \\ j \notin \cR }}\sum_{\substack{k\in[\sf K]\backslash \cR \colon \\ k\neq j} } \mat{H}_{j,k}\mat{U}_{\cR}\vect{v}_{\cR, k} \nonumber \\
&&+ \sum_{\substack{\cR \in [[\sf K]]^{\sf r}\colon \\ 1\in \cR, j \notin \cR }}\sum_{\substack{k\in[\sf K]\backslash \cR}}  \mat{H}_{j,k}\mat{U}_{\cR \cup \{k\}\backslash \{1\}}\vect{v}_{\cR, k} + \vect{Z}_j,\nonumber \\[-2ex]
&&\hspace{4.9cm}\quad j\in[\sf{K}]\backslash\{1\},\label{eq:Yj} \IEEEeqnarraynumspace
\end{IEEEeqnarray}
\end{subequations}
where for ease of notation we defined for Nodes $k\in[{\sf K}-1]$:
\begin{IEEEeqnarray}{rCl}
	\vect{v}_{\cR, k} &\triangleq&  \sum_{ \substack{j \in \cR}}  \vect{b}_{k,\cR \cup \{k\} \backslash\{j\}}^{j}, \qquad \forall \cR \in [[\sf K]\backslash \{k\}]^{\sf r},\label{eq:v_1}
	\end{IEEEeqnarray}
	and for the last Node $\sf K$, since its signal to Node 1 is absent: 
	\begin{IEEEeqnarray}{rCl}
	\vect{v}_{\cR, \sf K} &\triangleq& \sum_{ \substack{j \in \cR\backslash\{1\}}}  \vect{b}_{k,\cR \cup \{k\} \backslash\{j\}}^{j},  \qquad \forall \cR \in [[\sf K-1]]^{\sf r}. \label{eq:v_K}\IEEEeqnarraynumspace
\end{IEEEeqnarray}

Each Node $j$ zero-forces the non-desired interference terms of its ``cleaned" signal and  decodes  its intended messages.

\textit{Choice of IA Matrices $\{\mat{U}_{\cR}\}$:} 
Inspired by the IA scheme in \cite{jafar_degrees_2008}, we choose each $\mu \times \eta^\Gamma$ precoding matrix $\mat{U}_{\cR}$ so that its column-span includes all power products (powers  1 to $\eta$) of the channel matrices $\mat{H}_{j,k}$ that premultiply $\mat{U}_{\cR}$ in \eqref{eq:cleaned} in the non-desired interference terms. Thus,   $\cR \in [[\sf K]\backslash\{1\}]^{\sf r}$:
\begin{IEEEeqnarray}{rCl}\label{eq:U_cR}
\mat{U}_{\cR}\triangleq  \left[ \prod_{\mat{H} \in {\mathcal{H}}_{\cR} }
\mat{H}^{{\alpha_{\cR,\mat{H}}}}\cdot  \boldsymbol{\Xi}_{\cR}\colon  \,\; 
\forall \boldsymbol{\alpha}_{\cR} \in  [\eta]^{\Gamma} \right], \quad 
\end{IEEEeqnarray}
where $\{\boldsymbol{\Xi}_{\cR}\}_{\cR \in [[\sf K]\backslash\{1\}]^{\sf r}}$ are i.i.d. random vectors independent of all channel matrices, noises, and messages, 
\begin{IEEEeqnarray}{rCl}
	\label{eq:set_G}
	\mathcal{H}_{\cR} & \triangleq& \big\{\mat{H}_{j,k} \colon \;  j \in [\sf K]\backslash \cR, \ k \in [\sf K]\backslash  \{j\} \big\} \backslash \big\{ \mat{H}_{1,k} \colon k \in \cR\big\},\nonumber 
\end{IEEEeqnarray}
and $\boldsymbol{\alpha}_{\cR} \triangleq (\alpha_{\cR, \mat{H}}\colon \mat{H} \in {\mathcal{H}}_{\cR})$. Notice that $|\mathcal{H}_{\cR}|=\Gamma$ for any $\cR \in [[\sf K]\backslash\{1\}]^{\sf r}$.

\textit{Performance Analysis:} \sh{See the long version \cite{bi2022bounds}.}\lo{See Appendix~\ref{app:analysis}.}
\section{New Bounds on the NDT}\label{sec:main}


Define for any integer value $\sf r \in [\sf K]$:
\begin{IEEEeqnarray}{rCl} \label{eq:ndt_lower_bound}
     \Delta_{\textnormal{Ub}}(\sf r) \triangleq \left\{\begin{array}{ll}	\left(1-\frac{\sf r}{\sf K}\right) \cdot \frac{\sf r(\sf K-1) + \sf K - \sf r - 1}{\sf r(\sf K-1)^2 + \sf r(\sf K -2 )}   & \textrm{if }   \sf r < \sf K/2 \\[1.2ex]
  \frac{1}{\sf K}     \left(1 - \frac{\sf r}{\sf K}\right)& \textrm{if }  \sf  r  \geq   \sf K/2 \end{array} \right. . \IEEEeqnarraynumspace
\end{IEEEeqnarray} 
Also, let \begin{IEEEeqnarray}{rCl} \label{eq:ndt_upper_bound}
	\Delta_{\textnormal{Lb}}(\sf r) \triangleq \left\{
	\begin{aligned}
		&\frac{1}{\sf K}  \left(2 - \frac{3}{\sf K} \right)  &&\text{if } \sf r =1,\\
		&  \frac{1}{\sf K}  \left( 1 - \frac{\sf r}{\sf K}  + \!\max_{t \in[\lfloor \sf K/2 \rfloor] }\textnormal{lowc}\left( C_t( \sf r) \right)\right)&& \text{if } \sf r \in(1,2) ,\\
		&\frac{1}{\sf K}  \left( 1 - \frac{\sf r}{\sf K} +\textnormal{lowc}\left({C}_{\lfloor \sf K/2\rfloor}( \sf r ) \right)\right)  &&\text{if }\sf  r \in [2,\sf K],
	\end{aligned}
	\right.\nonumber\\
\end{IEEEeqnarray}
where for any $t \in [\lfloor \sf K/2\rfloor]$:
\begin{IEEEeqnarray}{rCl} \label{eq:C_i}
 C_t(i) &=& \begin{cases} \frac{\binom{\sf K - i}{t - i}}{\binom{\sf K}{t} \cdot t}\cdot (\sf K-t - i), &\textnormal{if  } i\in[t] ,\\
0,  & \textnormal{if }  i \in [\sf K]\backslash [t]
\end{cases} \IEEEeqnarraynumspace
\end{IEEEeqnarray}
and for any function $f$,  $\textnormal{lowc}\left(f(\ell)\right)$ denotes the lower convex  enveloppe of $\{ ( \ell, f(\ell))\}$.

\begin{theorem}\label{thm:ndt_bounds}
The NDT-computation  tradeoff $\Delta^*(\sf r)$ is upper-  and lower-bounded as:
\begin{IEEEeqnarray}{rCl}
	\Delta_{\textnormal{Lb}}(\sf r) \leq \Delta^*(\sf r) \leq  \textnormal{lowc}\left({\Delta}_{\textnormal{Ub}}(\sf r)\right).
\end{IEEEeqnarray}

\end{theorem}

\begin{IEEEproof}
For integers $\r \geq \K/2$ achievability of  $\Delta_{\textnormal{Ub}}(\sf r)$ is proved in \cite{li_wireless_2019}. For integers $\r < \K/2$ achievability of $\Delta_{\textnormal{Ub}}(\sf r)$ holds by Lemma~\ref{lem:E} and the  scheme  in Section \ref{sec:scheme}. Achievability of the lower convex enveloppe follows  by simple time- and memory-sharing strategies. The lower bound can be proved using MAC-type arguments, \sh{see our long version \cite{bi2022bounds}}\lo{see Appendix~\ref{app:converse}}. 
\end{IEEEproof}

\begin{remark}
The  upper bound is convex and piece-wise constant. 
 The lower bound is piecewise constant with segments spanning the   intervals $[i, i+1]$, for $i=2, \ldots, \K-1$. On the interval $[1,2)$, the lower bound is constant over smaller sub-intervals only but not over the entire segment.
 \end{remark}


\begin{corollary}\label{cor1}
For all $\sf r \geq \lceil \sf K/2 \rceil$, the linear interference cancelation scheme in \cite{li_wireless_2019} achieves the NDT, which equals
\begin{equation}\label{eq:cor}
\Delta^*(\sf r) = \left(1-\frac{\sf r}{\sf K}\right) \cdot \frac{1}{\sf K}.
\end{equation}
\end{corollary}
\begin{IEEEproof}
For $\sf r\geq  \lceil \sf K/2 \rceil$ the upper bound $ \text{lowc}\left(\Delta_{\textnormal{Ub}}(\sf r)\right)$ is equal to the lower bound $\Delta_{\textnormal{Lb}}(\sf r)$ because $C_{\lfloor \sf K/2 \rfloor}(i)=0$ for all $i\geq \lceil \sf K/2 \rceil$. 
\end{IEEEproof}

\begin{remark}\label{rem:better}
By \cite{li_wireless_2019}, $\Delta^*(\r)$  in \eqref{eq:cor} is achieved with beam-forming, zero-forcing, and side-information cancellation.  By Corollaries~\ref{cor1} and \ref{cor2},   these simple strategies are  sufficient to achieve $\Delta^*(\sf r)$ when $\sf r \geq \lceil \sf K/2 \rceil$ but not when $\sf r < \left\lceil \frac{\K-1}{2} \right \rceil$. 
\end{remark}

We compare the upper bound in Theorem \ref{thm:ndt_bounds} to  the bounds  in \cite{li_wireless_2019}  and \cite{bi_dof_2022}.  The upper bound in \cite{li_wireless_2019} is given as follows:
\begin{IEEEeqnarray}{rCl}\label{eq:LZF}
	\Delta^*(\sf r) \leq \Delta_{\textnormal{UB-BF}}(\r)\triangleq \text{lowc}\left\{ \left(\sf r, \frac{1 - \sf r/\sf K}{\min(\sf K, 2\sf r)}\right)\colon \sf r \in [\sf K] \right\}. \IEEEeqnarraynumspace
\end{IEEEeqnarray}
The upper bound in \cite{bi_dof_2022} has the  form: 
\begin{IEEEeqnarray}{rCl}
 	\lefteqn{\Delta^*(\sf  r)  \leq \Delta_{\textnormal{Ub-Groups}}(\r)\triangleq  } \nonumber \\
 	&& \textnormal{lowc}\left( (\mathsf{K}, 0)\cup \left\{ \left(\sf r, \, \frac{1-\sf r/{\sf K} }{\SDoFlb (\r)} \right)\colon \sf 1 \leq \sf r <\sf K, \sf r | \sf  K\right\}\right), \nonumber \\\label{eq:ISIT}
\end{IEEEeqnarray}
where 
\begin{IEEEeqnarray}{rCl} 
	\SDoFlb (\r)\triangleq  \left\{\begin{array}{ll} 2\sf r & \textrm{if }  \sf K/\sf r  \in \{2,3\}, \\  \frac{\sf K(\sf K-\sf r)-\sf r^2}{2\sf K-3\sf r} & \textrm{if }   \sf K/\sf r  \geq 4.\end{array} \right.
\end{IEEEeqnarray} 
Notice that  
 ${\Delta}_{\textnormal{Ub}}(1)= \Delta_{\textnormal{Ub-Groups}} (1)$. 

\begin{figure}[h!]
\vspace{-1mm}
\centering
	\scalebox{.82}{
		\begin{tikzpicture}[spy using outlines=
			{circle, lens={xscale=4, yscale=4}, connect spies}]
			\begin{axis}[
				xmin=1, xmax=11, xlabel={Computation Load ($\sf r$)},
				ymin=0, ymax=0.25, ylabel={NDT $\Delta^*(\sf r)$}]
							
				\addplot [color=red, line width=1, dashed] table [x index=0, y index=1]{figures/res_figure_data11/ONE_SHOT11.txt};
				\addlegendentry{
				One-shot scheme of \cite{li_wireless_2019}}
				\addplot [color=blue, line width=1, dashed] table [x index=0, y index=1]{figures/res_figure_data11/IA_ISIT11.txt};
				\addlegendentry{
				Grouped IA scheme of \cite{bi_dof_2022}}
			\addplot [color=green, line width=1] table [x index=0, y index=1]{figures/res_figure_data11/IA11.txt};
			\addlegendentry{
			Novel IA scheme}
				\addplot [line width=1, dotted] table [x index=0, y index=1]{figures/res_figure_data11/CONVERSE11.txt};
				\addlegendentry{Converse}
			\end{axis}
	\end{tikzpicture}
	}
	\vspace{-2mm}
	
	\caption{Bounds on $\Delta^*(\sf r)$  for $\K=11$.}
		\vspace{-2.5mm}
		
	\label{fig:comparison}
\end{figure}
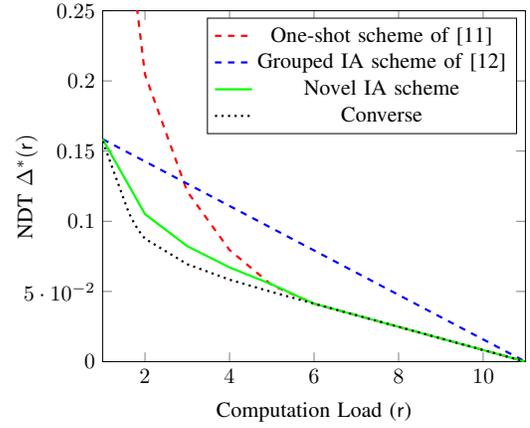
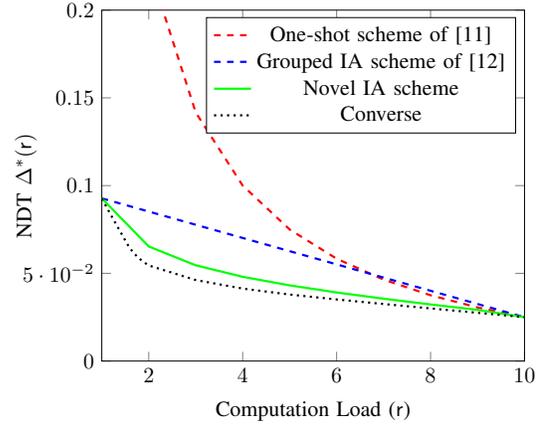
\begin{figure}[h!]
	
	\centering
	\scalebox{.82}{
		\begin{tikzpicture}[spy using outlines=
			{circle, lens={xscale=10, yscale=4}, connect spies}]
			\begin{axis}[
				xmin=1, xmax=10, xlabel={Computation Load ($\sf r$)},
				ymin=0, ymax=0.2, ylabel={NDT $\Delta^*(\sf r)$},
				]
					\addplot [color=red, line width=1, dashed] table [x index=0, y index=1]{figures/res_figure_data20/ONE_SHOT20.txt};
				\addlegendentry{One-shot scheme of \cite{li_wireless_2019}}
				\addplot [color=blue, line width=1, dashed] table [x index=0, y index=1]{figures/res_figure_data20/IA_ISIT20.txt};
				\addlegendentry{Grouped IA scheme of \cite{bi_dof_2022}}
				\addplot [color=green, line width=1] table [x index=0, y index=1]{figures/res_figure_data20/IA20.txt};
				\addlegendentry{Novel IA scheme}
				\addplot [line width=1, dotted] table [x index=0, y index=1]{figures/res_figure_data20/CONVERSE20.txt};
				\addlegendentry{Converse}
			\end{axis}
		\end{tikzpicture}
	}
			\vspace{-2mm}

	\caption{Bounds on $\Delta^*(\sf r)$  for $\sf K = 20$.}
	\label{fig:comparison2}
	\vspace{-3mm}
	
\end{figure}

\begin{corollary}\label{cor2}
For all $1<\r <  \left \lceil  \frac{\K}{2} \right \rceil$:
\begin{equation}\label{eq:ineqGroups}
\Delta^*(\sf  r)  \leq \textnormal{lowc}\left({\Delta}_{\textnormal{Ub}}(\sf r)\right) < \Delta_{\textnormal{Ub-Groups}} (\r),
\end{equation} 
and for all $1\leq \r <  \left \lceil  \frac{\K-1}{2} \right \rceil$:
\begin{equation}\label{eq:ineqZF}
\Delta^*(\sf  r)  \leq \textnormal{lowc}\left({\Delta}_{\textnormal{Ub}}(\sf r)\right) < \Delta_{\textnormal{Ub-ZF}} (\r).
\end{equation} 
\end{corollary} 
Fig.~\ref{fig:comparison} and \ref{fig:comparison2} compare the bounds in Theorem~\ref{thm:ndt_bounds} to the previous upper bounds $ \Delta_{\textnormal{Ub-Groups}} (\r)$ and $\Delta_{\textnormal{UB-BF}} (\r)$.

\section{Conclusion}
This paper presents an improved upper bound  and the first information-theoretic lower bound on the NDT tradeoff of full-duplex wireless MapReduce systems. The upper bound is obtained by  zero-forcing and a novel IA scheme that is tailored to the information cancellation capabilities of the nodes in a MapReduce system. As a conclusion of this work, we observe that linear beamforming, zero-forcing, and interference cancelation are optimal when each node can store at least half of the file, but they are suboptimal for smaller computation loads.

\section*{Acknowledgements} We thank P. Ciblat for helpful discussions. This work has been supported by National Key R\&D Program of China under Grant No 2020YFB1807504 and National Science Foundation of China Key Project under
Grant No 61831007.

\bibliographystyle{IEEEtran}
\bibliography{IEEEabrv,references}
\clearpage
\appendices

\lo{
\section{Proof of Lemma~\ref{lem:E}}\label{app:lem1}
We  show how to construct a distributed computing scheme achieving the NDT upper bound in \eqref{eq:DSDof}.  We shall assume a sequence (in $\sf P>0$) of rates $( R_{\mathcal{T}}^j \colon \mathcal{T}\in [\K]^\r, \; j\in [\K]\backslash \mathcal{T})$ that  achieves the sum-DoF $\SDoF(\r)$ and is completely symmetric with respect to indices $j$ and sets $\mathcal{T}$. By the same time-sharing and relabeling arguments as described in Subsection~\ref{sec:WDC_system} such a sequence must exist.

In the Map Phase we choose a regular file assignment. 
Partition the input files  $\{W_1, \ldots, W_{\sf N}\}$ into 
$\sf K \choose \sf r$ disjoint bundles  and assign each bundle to a size-$\sf r$ subset $\cT\in [[\sf K]]^{\sf r}$. The bundle associated to subset $\cT$ is denoted $\mathcal{W}_{\cT} \subseteq \{W_1, \ldots, W_{\sf N}\}$.
Notice that the proposed assignment   satisfies the constraint on the computation load, because the number of files stored at Node $k$ is: 
\begin{equation}
\binom{\sf K-1}{\sf r-1}\frac{\sf N}{\binom{\sf K}{\sf r}}= \frac{\sf r}{\sf K} \sf N.
\end{equation}

Each node computes all IVAs associated to its stored files. 

During the Wireless Shuffle Phase, each transmit set $\mathcal{T}$ communicates to any receive node $j\notin \mathcal{T}$ all IVAs that can be calculated from bundle  $\mathcal{W}_{\cT}$. To this end, all nodes use the encoding and decoding functions  achieving  $\SDoF(\r)$.

Each transmit group $\mathcal{T}$ has to send $\frac{\sf N}{{\K \choose \r}}$ IVAs to each Node $j\notin \mathcal{T}$ and in total there are ${ \K \choose \r} (\K-\r)$ rates in the symmetric rate vector $( R_{\mathcal{T}}^j \colon \mathcal{T}\in [\K]^\r, \; j\in [\K]\backslash \mathcal{T})$. By definition, the  probability of error in reconstructing all missing IVAs  tends to 0 as $\sf T \to \infty$ if 
\begin{equation}\label{eq:Tprime}
\varlimsup_{P\to \infty}  \varlimsup_{A \to \infty} \frac{\sf A \cdot  \sf N}{\sf T \cdot {\K \choose \r}\cdot \log \sf P } <   \frac{ \SDoF(\r)}{{ \K \choose \r}(\K-\r)}.
\end{equation}
We thus conclude that 
\begin{equation}\label{eq:Tprime}
\Delta^\star(\r) \leq \varliminf_{P\to \infty}  \varliminf_{A \to \infty} \frac{\sf T}{\sf A \cdot  \sf K\cdot \sf N} \cdot \log \sf P=  \frac{\K-\r }{\K }      \frac{ 1}{\SDoF(\r)},
\end{equation}
 which proves the desired achievability result.
}

\lo{
\section{Analysis of IA Scheme}\label{app:analysis}

By the choice of $\sf T$ in \eqref{eq:T},  the signal and interference space at  Rx $1$ is represented by the $ \mu \times  \mu$-matrix:
\begin{IEEEeqnarray}{l}\label{eq:signal_interference1}
\boldsymbol{\Lambda}_1= \Big[ \underbrace{ \mat{D}_1 }_{\textnormal{signal space}}, \; \underbrace{ \left[\mat{W}_{\cR}\right]_{\substack{\cR\in [[\sf K]\backslash\{1\}]^{\sf r}}}}_{\textnormal{interference space}} \Big]
, \IEEEeqnarraynumspace
\end{IEEEeqnarray}
where the signal subspace is given by a collection of  $\sf \mu \times \left((\sf K -2) \cdot \binom{\sf K - 2}{\sf r -1} \cdot \eta^{\Gamma}\right)$-matrices
\begin{IEEEeqnarray}{rCl} \label{eq:D_1}
    \mat{D}_{1} \triangleq \Big[\mat{H}_{1,k}\mat{U}_{\cR}\Big]_{\substack{k \in [\sf K-1]\backslash\{1\} , \\ \cR\in [[\sf K]\backslash\{1\}]^{\sf r}\colon\\  k \in \cR }}.
\end{IEEEeqnarray}
The signal space at Rx~$j \in [\sf K] \backslash\{1\}$ is represented by the $\mu \times \tilde{\mu}$ matrix: 
\begin{IEEEeqnarray}{rCl}\label{eq:signal_interference2}
	\boldsymbol{\Lambda}_j \triangleq
	&&  \Big[ \underbrace{ \mat{D}_j }_{\textnormal{signal space}},  \; \underbrace{ \left[\mat{W}_{\cR}\right]_{\substack{\cR\in [[\sf K]\backslash\{1\}]^{\sf r}\colon j \notin \cR}}}_{\textnormal{interference space}} \Big] .  \IEEEeqnarraynumspace
\end{IEEEeqnarray}
where the signal subspace $\mat{D}_j$ is  given by the collection of $ \mu \times \left(\sf r \cdot \binom{\sf K -1}{\sf r} \cdot \eta^{\Gamma}\right)$-matrices
\begin{IEEEeqnarray}{rCl} \label{eq:D_j}
	\mat{D}_j  \triangleq \left[\mat{H}_{j,k}\mat{U}_{\cR}\right]_{\substack{\cR\in [[\sf K]\backslash\{1\}]^{\sf r}\colon \\  j \in \cR, \\k \in [\sf K]\backslash\{j\}}}
\end{IEEEeqnarray}
and 
\begin{IEEEeqnarray}{rCl}
	\tilde{\mu}\triangleq \sf r \cdot \binom{\sf K-1}{\sf r} \cdot \eta^{\Gamma} + \binom{\sf K-2}{\sf r}\cdot (\eta+1)^\Gamma.
\end{IEEEeqnarray}

According to Lemmas~\ref{lemma1} and \ref{lma:diag_fullrank} below, $\{\mat{\Lambda}_j\}$ is full column-rank if each column has different exponent vector $\boldsymbol{\alpha}$, which follows by the way we constructed the matrices $\mat{U}_{\cR}$ and $\mat{W}_{\cR}$. Indeed:
\begin{itemize}
	\item For each $\cR \in [[\sf K]\backslash\{1\}]^{\sf r}$, matrices $\mat{U}_{\cR}$ and $\mat{W}_{\cR}$ are constructed using a dedicated i.i.d. vector $\boldsymbol{\Xi}_{\cR}$ that is independent of all other random variables in the system and thus the vectors $\boldsymbol{\Xi}_{\cR}$ can play the roles of the vectors $\boldsymbol{\Xi}_{i}$ in Lemma~\ref{lma:diag_fullrank}.
	\item For each term $\mat{H} \mat{U}_{\cR}$ in \eqref{eq:D_1} and \eqref{eq:D_j}, we have $\mat{H} \notin \mathcal{H}_{\cR}$. Thus $\mat{H}$ is not used in the construction of neither $\mat{U}_{\cR}$ nor $\mat{W}_{\cR}$ and induces a unique exponent on the corresponding columns in the signal space which is 0 in all columns of the interference space $\mat{W}_{\cR}$.
\end{itemize}
This proves that based on the ``cleaned" signal \eqref{eq:cleaned}, each receiving node $j$ can separate the various desired   signals from each other as well as from the non-desired interfering signals. 
Since each codeword $\vect{b}_{k,\mathcal{T}}^j$ occupies $\eta^\Gamma$ dimensions out of the $\mu$ dimensions, we obtain that whenever
\begin{IEEEeqnarray}{rCl}\label{eq:rel}
\frac{|\vect{b}_{k,\mathcal{T}}^j| }{ \sf T} \leq  \frac{\eta^\Gamma}{ \sf T} \log \sf P + o(\log \sf P),
\end{IEEEeqnarray}
for an appropriate function $o(\log \sf P)$  that grows slowlier than $\log \sf P$, each codeword  $\vect{b}_{k,\mathcal{T}}^j$  can be decoded with arbitrary small probability of error as $ \eta \to \infty$.

Since $(\K-2)\cdot { \K -2 \choose \r-1}$ codewords are sent to Node 1,  and $\r {\K-1 \choose \r}$ codewords to any other Node $j=2,\ldots,\K$, and since 
\begin{equation}
\lim_{\eta \to \infty}\frac{\eta^\Gamma}{ \sf T}  = \frac{1}{ {(\K-2) \binom{\sf K -2}{\sf r-1}} +{ \binom{\sf K-1}{\sf r}}},
\end{equation} we conclude that a sum-DoF of 
\begin{IEEEeqnarray}{rCl}
\SDoF& = &\frac{ (\K-2)\cdot { \K -2 \choose \r-1} + (\K-1) {\r {\K-1 \choose \r} }}{ (\K-2){\binom{\sf K -2}{\sf r-1 }}+{ \binom{\sf K-1}{\sf r}} }\nonumber\\
& =& \frac{ \r (\K-1)^2 +\r (\K-2)}{\r (\K-2)+\K-1}
\end{IEEEeqnarray}
is achievable over the system. This establishes the desired achievability result. 

\begin{lemma}\label{lemma1}
Let $\*s_1, \*s_2, ..., \*s_m$ be  independent random vectors with i.i.d. entries drawn according to continuous distributions.  
for any $L\leq m$ and $L$ \emph{different} exponent vectors 
\begin{equation*}
    \boldsymbol{\alpha}_j = \left(\alpha_{j,1}, \ldots, \alpha_{j,m}\right) \in \mathbb{Z}^m_{+}, \quad  j \in [L],
\end{equation*}
the  $m \times L$ matrix $\*M$ with row-$i$ and column-$j$ entry
\begin{equation}
    M_{i,j}= \prod_{k=1}^m
    \left(s_{i,}\right)^{\alpha_{j,k}}, \quad 
    i \in [m],\; j \in [L],
\end{equation}
 is full rank almost surely.
\end{lemma}

\begin{lemma} \label{lma:diag_fullrank}
Consider numbers $\{n_1, n_2, \cdots, n_{\tilde{\sf K}}\} \in \mathbb{Z}_{+}^{\tilde{\sf K}}$ so that their sum  $C \triangleq \sum_{i=1}^{\tilde{\sf K}} n_i \leq \mu$. Assume that for each $i \in [\tilde{\sf K}]$ and $k \in [n_i]$,   $\mat{B}_{i,k} \in \mathbb{C}^{\mu \times \mu}$ is a diagonal matrix so that all square sub-matrices of the following matrices  $\{\mat{B}_{i}\}_{i\in\tilde{\sf K}}$ are full rank:
\begin{equation}
    \mat{B}_{i} \triangleq \big[ \mat{B}_{i,1} \cdot \boldsymbol{1}_{\mu}, \mat{B}_{i,2} \cdot \boldsymbol{1}_{\mu}, \cdots, \mat{B}_{i,n_i} \cdot \boldsymbol{1}_{\mu} \big], \quad  i \in [\tilde{\sf K}],
\end{equation}
where $\boldsymbol{1}_{\mu}$ denotes a $\mu$-dimensional all-one column vector.

Let further  $\{\boldsymbol{\Xi}_i\}_{i \in \tilde{\sf K}}$ be independent $\mu$-vectors with entries drawn i.i.d. from continuous distributions and define the $\mu\times n_i$-matrices
\begin{equation}
    \mat{A}_i \triangleq \left[ \mat{B}_{i,1} \cdot \boldsymbol{\Xi}_{i}, \mat{B}_{i,2} \cdot \boldsymbol{\Xi}_{i}, \cdots, \mat{B}_{i,n_i} \cdot \boldsymbol{\Xi}_{i} \right], \quad i \in [\tilde{\sf K}].
\end{equation}

Then, the  $\mu \times C$-matrix 
\begin{equation}
    \mat{\Lambda} \triangleq \left[ \vect{A}_1, \vect{A}_2, \cdots, \vect{A}_{\tilde{\sf K}} \right]
\end{equation}
has full column rank almost surely. 
\end{lemma}
\begin{IEEEproof}
We assume that the matrix $\*\Lambda$ is a square matrix i.e. $C = \mu$. If $\mu > C$, we take a square submatrix of $\* \Lambda$ and  perform the same proof steps on the submatrix.

Define
\begin{equation}
    F \left( \boldsymbol{\Xi}_1, \ldots, \boldsymbol{\Xi}_{\tilde{\sf K}} \right) \triangleq \det (\*\Lambda)
\end{equation}
which is a polynomial of $\boldsymbol{\Xi}_1, \boldsymbol{\Xi}_2, \cdots, \boldsymbol{\Xi}_{\tilde{\sf K}}$ as the determinant is a polynomial of the entries of $\mat{\Lambda}$.

For the vectors
\begin{IEEEeqnarray}{rCl}
\boldsymbol{d}_i =[\underbrace{0, \cdots 0,}_{ (n_1+\cdots +n_{i-1})\textnormal{ 0s}} \underbrace{1, \cdots 1, }_{n_i \textnormal{ 1s}} \underbrace{0, \cdots 0}_{ (n_{i+1}+\cdots +n_{\tilde{\sf K}})\textnormal{ 0s}}]^\intercal , \quad i\in \tilde{\sf {K}}, \IEEEeqnarraynumspace
\end{IEEEeqnarray}
the polynomial evaluates to
\begin{IEEEeqnarray}{rCl}
F\left( \boldsymbol{d}_1,  \ldots, \boldsymbol{d}_{\tilde{\sf K}} \right) &=& \det 
\begin{pmatrix}
\mat{B}_1' & \mat{0} & \cdots & \mat{0}\\
\mat{0} & \mat{B}_2' & \cdots & \mat{0}\\
\vdots & \vdots & \ddots & \vdots \\
\mat{0} & \mat{0} & \cdots & \mat{B}_{\tilde{\sf K}}'
\end{pmatrix} \\
&=& \prod_{i=1}^{\tilde{\sf K}} \det(\mat{B}_i') \neq 0
\end{IEEEeqnarray}
where $\mat{B}'_i$ is the $n_i \times n_i$ square sub-matrix of $\mat{B}_i$ consisting  of its rows $(n_1+\cdots+n_{i-1}+1)$ to $(n_1+\cdots+n_{i-1}+n_i)$. The inequality holds by our assumption that all square sub-matrices of $\mat{B}_i$ are full rank. 

We conclude that $F$ is a non-zero polynomial and thus $F \left( \boldsymbol{\Xi}_1, \ldots, \boldsymbol{\Xi}_{\tilde{\sf K}} \right) $ equals $0$ with probability $0$ because the entries of $\boldsymbol{\Xi}_1, \boldsymbol{\Xi}_2, \cdots, \boldsymbol{\Xi}_{\tilde{\sf K}}$ are drawn independently from continuous distributions.
\end{IEEEproof}
}

\lo{
\section{Proof of the NDT Lower Bound in Theorem~\ref{thm:ndt_bounds}}
\label{app:converse}
Consider a fixed file assignment (map phase), 
and for any positive power $\sf P$ a sequence (in $\sf T$) of wireless distributed computing systems satisfying \eqref{eq:error_computing} for the given file assignment.  (Since for finite $\sf N$ there are only a finite number of different file assignments irrespective of $\sf P$ and $\sf T$, we can fix the assignment.)
The following limiting behaviour must hold. 
\begin{lemma}\label{lma:dof_upper_bound}
	Consider two disjoint sets $\cT$ and $\cR$ of same size 
	\begin{equation}
		|\cT| = |\cR|,
	\end{equation}
	and define $\mathcal{F}\triangleq[\sf K]\backslash (\mathcal{R} \cup\mathcal{T})$. Let $\mathcal{M} \subseteq [\sf N]$ be the set of  files known only to nodes $\cT$ but not to any other node and partition the set of all IVAs $\mathcal{A}$ it into the following disjoint subsets:
	\begin{IEEEeqnarray}{rCl}
		\mathcal{W}_r & \triangleq & \{a_{j,m}\}_{\substack{j \in \mathcal{R}\\m \in [\sf N]\backslash\mathcal{M}_j}}, \\
		\mathcal{W}_t & \triangleq  & \{a_{j,m}\}_{\substack{j \in (\mathcal{T}\cup\mathcal{F}) \\ m \in \mathcal{M}\backslash\mathcal{M}_j}}.
	\end{IEEEeqnarray}
		For any sequence of distributed computing systems:
	\begin{equation}\label{eqn:thm_1}
				d \triangleq \varlimsup_{P \to \infty}  \varlimsup_{\sf T \to \infty}\frac{ \sf A  }{ \sf T \log \sf P} \leq\frac{ |\mathcal{T}|}{|\mathcal{W}_t| + |\mathcal{W}_r|}
	\end{equation}
\end{lemma}
(Notice that $\mathcal{W}_r$ denotes the set of all IVAs intended to nodes in $\mathcal{R}$ and $\mathcal{W}_t$ the set of  IVAs deduced from files in $\mathcal{M}$ and intended for nodes not in $\mathcal{R}$.)
\begin{proof}
		Denote by $\mathcal{H}$ the set of all channel coefficients to all nodes in the system and define $\mathcal{W}_c  \triangleq  \mathcal{A} \backslash (	\mathcal{W}_r  \cup 	\mathcal{W}_t$. 
	Since channel coefficients and IVAs are independent, we have
	\begin{IEEEeqnarray}{rCl}
		\lefteqn{	H(\mathcal{W}_t, \mathcal{W}_r) }\nonumber \\
		&=& H(\mathcal{W}_t, \mathcal{W}_r|\mathcal{W}_c,\mathcal{H}) \\
		&=& I(\mathcal{W}_t, \mathcal{W}_r; \*Y_{\mathcal{R}}|\mathcal{W}_c,\mathcal{H}) + 
		H(\mathcal{W}_t, \mathcal{W}_r| \mathcal{W}_c,\*Y_{\mathcal{R}}, \mathcal{H}) \IEEEeqnarraynumspace\\
		&=& h(\*Y_{\mathcal{R}}| \mathcal{W}_c,\mathcal{H}) - h(\*Z_{\mathcal{R}})  \nonumber \\ 
		&&+ H(\mathcal{W}_r| \mathcal{W}_c,\*Y_{\mathcal{R}}, \mathcal{H})
		+ H(\mathcal{W}_t|\mathcal{W}_r,\mathcal{W}_c, \*Y_{\mathcal{R}}, \mathcal{H}) \\
		&\leq & h(\*Y_{\mathcal{R}}|\mathcal{W}_c, \mathcal{H}) - h(\*Z_{\mathcal{R}}) \nonumber \\
		&& +  {\sf T} \epsilon_{\sf T}
		+ H(\mathcal{W}_t|\mathcal{W}_r,\mathcal{W}_c, \*Y_{\mathcal{R}}, \mathcal{H}) ,\label{eq:a2}
	\end{IEEEeqnarray}
	where we defined $\*Y_{\mathcal{A}} \triangleq [\*Y_j]_{j \in \mathcal{A}}$ for a set $\mathcal{A}\subseteq [\sf K]$ and $\epsilon_{\sf T}$ is a vanishing sequence  as $\sf T\to \infty$. Here the  inequality holds by Fano's inequality, because $\mathcal{W}_r$ is decoded from  $\*Y_{\mathcal{R}}$ and $\mathcal{W}_c$, and because we impose vanishing probability of error \eqref{eq:error_computing}. 
		
	Again by Fano's inequality and by \eqref{eq:error_computing}, there exists a vanishing sequence $\epsilon_{\sf T}'$ such  that
	\begin{IEEEeqnarray}{rCl} 
		\lefteqn{H(\mathcal{W}_t|\mathcal{W}_r,\mathcal{W}_c, \*Y_{\mathcal{R}},\mathcal{H}) }\hspace{1cm}  \nonumber \\
		&\leq & I (\mathcal{W}_t; \*Y_{(\mathcal{F} \cup \mathcal{T})}|\mathcal{W}_r,\mathcal{W}_c, \*Y_{\mathcal{R}}, \mathcal{H}) 
		+ \sf{T} \epsilon_{\sf T}' \\
		&=& h(\*Y_{(\mathcal{F}\cup \mathcal{T})}| \mathcal{W}_r,\mathcal{W}_c, \*Y_{\mathcal{R}}, \mathcal{H}) \\
		&&- h(\*Y_{(\mathcal{F} \cup \mathcal{T})}| \mathcal{W}_r, \mathcal{W}_t,\mathcal{W}_c, \*Y_{\mathcal{R}},  \mathcal{H}) + \sf{T} \epsilon_{\sf T}' \\
		&\leq& h(\bar{\*Y}_{(\mathcal{F} \cup \mathcal{T})}|\bar{\*Y}_{\mathcal{R}}, \mathcal{H})   - h(\*Z_{(\mathcal{F} \cup \mathcal{T})}) + \sf{T} \epsilon_{\sf T}' , \IEEEeqnarraynumspace \label{eq:a}
	\end{IEEEeqnarray}
	where  $\bar{\*Y}_{\mathcal{A}} \triangleq [\bar{\*Y}_j]_{j \in \mathcal{A}}$ and $\bar{\*Y}_j$ denotes Node $j$'s ``cleaned" signal without the inputs that do not depend on files in $\mathcal{M}$ but   only  on IVAs $\mathcal{W}_r \cup  \mathcal{W}_c$: 
	\begin{equation*}
		\bar{\*Y}_j \triangleq \*H_{j,{\mathcal{T}}}\vect{X}_{\mathcal{T}} + \*Z_j, \qquad  j \in \mathcal{T}\cup \mathcal{F}.
	\end{equation*}
	Here,	$\*H_{\mathcal{A},\mathcal{B}}$ denotes the channel matrix from  set $\mathcal{B}$ to  set $\mathcal{A}$. 
	
	To bound the first term in \eqref{eq:a}, we introduce a random variable $E$ indicating whether the matrix $\mat{H}_{\mathcal{R},\mathcal{T}}$ is invertible $(E=1)$  or not $(E=0)$. If this matrix is invertible and $E=1$, then  the input vector  $\vect{X}_{\mathcal{T}}$ can be computed from  $\bar{\*Y}_{\mathcal{R}}$ up to noise terms.   Based on this observation and defining the residual noise terms
	\begin{equation}
		\bar{\*Z}_j \triangleq \*Z_j - \*H_{j,\mathcal{T}}\*H_{\mathcal{R},\mathcal{T}}^{-1}\*Z_{\mathcal{R}}, \quad \textnormal{ if }E=1,
	\end{equation}
	we obtain: 
	\begin{IEEEeqnarray}{rCl}
		\lefteqn{	h(\bar{\*Y}_{(\mathcal{F}\cup \mathcal{T})}| \bar{\*Y}_{\mathcal{R}}, \mathcal{H}) }\nonumber\\
		&\leq & \mathbb{P}(E=1) \cdot h\left(\bar{\mat{Z}}_{(\mathcal{F}\cup \mathcal{T})} |\bar{\*Y}_{\mathcal{R}},\mathcal{H}, E=1\right)  \nonumber\\
		&&+ \mathbb{P}(E=0) \cdot h\left(\bar{\*Y}_{(\mathcal{F} \cup \mathcal{T})} | \bar{\*Y}_{\mathcal{R}},\mathcal{H}, E=0\right) \\
		& \leq & h\left(\bar{\mat{Z}}_{(\mathcal{F}\cup \mathcal{T})}\right)  + \mathbb{P}(E=0)h\left(\bar{\*Y}_{(\mathcal{F} \cup \mathcal{T})} | \bar{\*Y}_{\mathcal{R}},\mathcal{H}, E=0\right).\IEEEeqnarraynumspace
	\end{IEEEeqnarray}
	Since  the channel coefficients follow continuous distribution, $\mat{H}_{\mathcal{R},\mathcal{T}}$ is invertible almost surely,  implying $\mathbb{P}(E=0)=0$. By the boundedness of the entropy term $h\left(\bar{\*Y}_{(\mathcal{F} \cup \mathcal{T})} | \bar{\*Y}_{\mathcal{R}},\mathcal{H}, E=0\right)$ (since power $\sf P$ and channel coefficients are bounded), this implies 
		\begin{IEEEeqnarray*}{rCl}
		h(\bar{\*Y}_{(\mathcal{F}\cup \mathcal{T})}| \bar{\*Y}_{\mathcal{R}}, \mathcal{H}) \leq   h(\bar{\mat{Z}}_{(\mathcal{F} \cup \mathcal{T})}),
	\end{IEEEeqnarray*}
	which combined with \eqref{eq:a2} and \eqref{eq:a} yields:
	\begin{IEEEeqnarray}{rCl} 
		H(\mathcal{W}_t, \mathcal{W}_r) 
		&\leq & h(\*Y_{\mathcal{R}}|\mathcal{H}) - h(\*Z_{\mathcal{R}})
		+ h(\bar{\*Z}_{(\mathcal{F} \cup \mathcal{T})}) \nonumber\\
		&&- h(\*Z_{(\mathcal{F}\cup \mathcal{T})}) +  {\sf T} (\epsilon_{\sf T} + \epsilon_{\sf T}') \nonumber\\
		&\leq& {\sf T}|\mathcal{R} |\log(\sf P) + {\sf T}C_{\sf T, \mathcal{H}}, \label{eqn:converse_part4_2}
	\end{IEEEeqnarray}
	where $C_{\sf T, \mathcal{H}}$ is a function that is uniformly bounded over all realizations of  channel matrices and powers $\sf P$. Noticing  
	\begin{equation}
		H(\mathcal{W}_t, \mathcal{W}_r) = \sf A( |\mathcal{W}_t|+| \mathcal{W}_r|),
	\end{equation} dividing  \eqref{eqn:converse_part4_2} by $\sf T \log (\sf P)$,  and letting $\sf P \to \infty$, establishes the lemma because $|\mathcal{R}|=|\mathcal{T}|$ and  ${\sf T}C_{\sf T, \mathcal{H}}$ is bounded. 
\end{proof}

For each subset $\mathcal{T}\subseteq [\sf K]$, let  $\mathcal{B}_{\mathcal{T}}^j$ denote the set of IVAs that are computed exclusively  at nodes in set $\mathcal{T}$ and intended for  reduce function $j$.  Define  $b_{\mathcal{T}} =|\mathcal{B}_{\mathcal{T}}^j|$, which does not depend on the index of the reduce function  $j \in [\sf K]\backslash\mathcal{T}$. 

Choose two disjoint subsets $\cT$ and $\cR$ of same size $|\cT|=|\cR|$. By  Lemma \ref{lma:dof_upper_bound}, and rewriting the sets $\mathcal{W}_t$ and $\mathcal{W}_r$ in the lemma in terms of the sets $\{\mathcal{B}_{\mathcal{T}}^j\}$,  we obtain: \begin{IEEEeqnarray}{rCl}
	\frac{|\cT|}{d} &\geq& \sum_{\mathcal{T}  \subseteq [\sf K]} \; \;\sum_{j\in \RM \backslash \mathcal{T}} |\mathcal{B}_{\mathcal{T}}^j| + \sum _{\mathcal{G} \subseteq \TM} \;\; \sum_{j \in [\sf K]\backslash (\RM \cup \mathcal{G})} |\mathcal{B}_{\mathcal{G}}^j|\IEEEeqnarraynumspace \\
	&=& \sum_{\mathcal{T} \subseteq [\sf K]} |\RM \backslash \mathcal{T}| \cdot b_{\mathcal{T}} + \sum _{\mathcal{G} \subseteq \TM} (\sf K - |\RM| - |\mathcal{G}|) \cdot b_{\mathcal{G}} . \label{eq:converse_constrain}
\end{IEEEeqnarray}

Summing up Equality \eqref{eq:converse_constrain} over all  sets  $\cT$ and $\cR$  of constant size $t\leq \sf K/2$, we obtain:
\begin{IEEEeqnarray}{rCl}
	\lefteqn{\binom{\sf K}{t} \cdot \binom{\sf K -t}{t} \cdot \frac{t}{d} } \quad \nonumber \\ 
	&\geq& \sum_{\TM \in[[\sf K]]^t}  \;\; \sum_{\RM \in [ [\sf K]\backslash \TM]^t}  \;\;\sum_{\mathcal{T} \subseteq [\sf K]} |\RM \backslash \mathcal{T}| \cdot b_{\mathcal{T}} \nonumber \\
	&&+ \sum_{\TM \in[[\sf K]]^t} \;\;  \sum_{\RM \in [ [\sf K]\backslash \TM]^t}  \;\; \sum _{\mathcal{G} \subseteq \TM} (\sf K -t - |\mathcal{G}|) \cdot b_{\mathcal{G}}  \IEEEeqnarraynumspace \\
		&=& \sum_{\mathcal{T} \subseteq[\sf K]}  \binom{\sf K}{t} \cdot \binom{\sf K -t}{t} \cdot(\sf K -|\mathcal{T}|) \frac{t}{\sf K} b_{\mathcal{T}} \nonumber\\
	&&+\sum_{\substack{\mathcal{G}\subseteq [\sf K] \colon\\ |\mathcal{G}|\leq t}} \binom{\sf K -|\mathcal{G}|}{t - |\mathcal{G}|} \cdot  \binom{\sf K -t}{t}\cdot (\sf K -t - |\mathcal{G}|) b_{\mathcal{G}} \\
	& = & \binom{\sf K}{t} \cdot \binom{\sf K -t}{t} \cdot t \left( \sf N  - \frac{ \sf r \sf N}{\sf K}\right) \nonumber \\
	&&+ \sum_{\substack{\mathcal{G}\subseteq [\sf K] \colon\\ |\mathcal{G}|\leq t}} \binom{\sf K -|\mathcal{G}|}{t - |\mathcal{G}|} \cdot \binom{\sf K -t}{t}\cdot (\sf K -t - |\mathcal{G}|) b_{\mathcal{G}} ,  \label{eq:converse_constrain_1} \IEEEeqnarraynumspace
\end{IEEEeqnarray}
where we define $\binom{a}{0}=1$ for any positive integer $a$. The first equality holds because for a given set $\mathcal{T}$, each element of $[\sf K]\backslash \mathcal{T}$ is present in a fraction of $t/{\sf K}$   pairs of  the admissible sets $(\mathcal{R},\mathcal{T})$ and the last equality holds because 
\begin{IEEEeqnarray}{rCl}
	\sum_{\mathcal{T} \subseteq [\sf K]} b_{\mathcal{T}} = \sf N, \quad \sum_{\mathcal{T} \subseteq [\sf K]} |\mathcal{T}| \cdot b_{\mathcal{T}} \leq \sf r \cdot \sf N. 
\end{IEEEeqnarray}
Dividing both sides of \eqref{eq:converse_constrain_1} by $\binom{\sf K}{t}  \binom{\sf K -t}{t} t$, and defining $b_i \triangleq \sum_{\substack{\mathcal{T}\in[ [\sf K]]^i}} b_{\mathcal{T}}$,
for any $t\in [\lfloor\sf K/2\rfloor]$ we obtain:
\begin{IEEEeqnarray}{rCl}\label{eq:min}
	\frac{1}{d}& \geq &\sf N - \frac{\sf r \cdot \sf N}{\sf K} +\min_{\substack{b_1,\ldots, b_{\sf K} \in \mathbb{Z}^+\colon \\ \sum_{i=1}^{\sf K} b_i= \sf N \\ \sum_{i=1}^{\sf K} i b_i \leq \sf r \sf N}}
	\sum_{i=1}^t C_t(i)b_i, \quad  t \in [ \lfloor\sf K/2 \rfloor], \IEEEeqnarraynumspace
\end{IEEEeqnarray}
where  $C_t(i)$ is defined in \eqref{eq:C_i}. 

For any $t\in  [ \lfloor\sf K/2 \rfloor]$, the sequence of coefficients $C_{t}(1), C_{t}(2), \ldots, C_t(t)$ is convex and non-increasing, \lo{see Appendix~\ref{app:convexC}}\sh{see \cite[Appendix~A]{arxiv}}. Based on this convexity, it can be shown (see \lo{Appendix~\ref{app:optimality}}\sh{\cite[Appendix~B]{arxiv}})   that for any $\sf r < t+1$ there exists a solution to the minimization problem in  \eqref{eq:min} putting only positive masses on $b_{\lfloor \sf r\rfloor}^*$ and $b_{\lceil \sf r\rceil}^*$ in the unique  way satisfying 
\begin{IEEEeqnarray}{rCl}
	b_{\lfloor \sf r\rfloor}^* + b_{\lceil \sf r\rceil}^* & = & \sf N\\
	\lfloor \sf r\rfloor b_{\lfloor \sf r\rfloor}^* +  \lceil \sf r\rceil  b_{\lceil \sf r\rceil}^* &= & \sf r \sf N .
\end{IEEEeqnarray}
For $\sf r \geq t+1$ an optimal solution consists of setting $b_{\lfloor \sf r \rfloor}^*=\sf N$, in which case the minimization in \eqref{eq:min} evaluates to 0.  

For $\sf r \geq 2$, the lower bound on the NDT in the theorem is then obtained by plugging these optimum values into bound \eqref{eq:min} for the choice $t= \lfloor \sf K/2 \rfloor$. For $\sf r=1$ we choose  $t=1$, and for $\sf r \in (1,2)$ we  maximize over the value of $t$.
}

\lo{
\section{Proof of Monotonicity and Convexity of values $C_{i}^{(t)}$}\label{app:convexC}

We shall prove monotonicity and convexity of the values 
\begin{IEEEeqnarray}{rCl}
D^{(t)}_i & \triangleq & \frac{\sf K !}{t !}  t \cdot C^{(t)}_i   \\
& = & \frac{ (\sf K - i)!}{(t- i)!} (\sf K - t-i) , \qquad i \in [t].
\end{IEEEeqnarray}

Notice that 
\begin{IEEEeqnarray}{rCl}
D^{(t)}_{i-1} & = & D^{(t)}_{i}  \frac{\sf K -i +1}{t- i +1} \cdot \frac{\sf K -t -i+1}{\sf K -t -i} \\
D^{(t)}_{i+1} & = & D^{(t)}_{i}  \frac{t- i }{\sf K -i} \cdot \frac{\sf K -t -i-1}{\sf K -t -i}, 
\end{IEEEeqnarray}
and the monotonicity  
\begin{equation}
D^{(t)}_{i-1} > D^{(t)}_{i}
\end{equation}
simply follows because  $K>t$ implies
\begin{equation}
 {\sf K -i +1} > {t -i+1} \quad \textnormal{and} \quad {\sf K- t- i +1} > {\sf K -t -i} .
\end{equation}

To prove convexity, we shall prove  that 
\begin{IEEEeqnarray}{rCl}
D^{(t)}_{i+1} +D^{(t)}_{i-1} & \geq  & 2 D^{(t)}_{i},
\end{IEEEeqnarray}
or equivalently
\begin{IEEEeqnarray}{rCl}
\frac{t-i}{\sf K-i} \cdot \frac{ \sf K -t -i -1}{\sf K -t -i} + \frac{\sf K-i +1}{t-i +1} \cdot \frac{\sf K- t -i +1}{\sf K- t-i} \geq 2. \IEEEeqnarraynumspace
\end{IEEEeqnarray}
Multiplying both sides with the denominators, we see that this condition is equivalent to
\begin{IEEEeqnarray}{rCl}
\lefteqn{(t-i)( \sf K -t -i -1)(t -i+1) } \hspace{1cm}   \nonumber \\
\lefteqn{+ (\sf K-i +1)(\sf K- t -i +1)(\sf K-i )} \qquad \;\nonumber \\
&  \geq  &2 (\sf K-i) (t-i+1)(\sf K -t -i) .
\end{IEEEeqnarray}
and after rearranging the terms:
\begin{IEEEeqnarray}{rCl}
\lefteqn{ (\sf K-t)(\sf K-i)   (\sf K - t - i) + (\sf K-i)(\sf K -i+1)} \nonumber \\
& \geq& (\sf K -t ) (t-i+1 ) (\sf K-t-i) +(t-i) (t-i+1). \IEEEeqnarraynumspace
\end{IEEEeqnarray}
This last inequality is easily verified by noting that  $\sf K \geq t+1$.

\section{Proof of Structure of Minimizer}\label{app:optimality}
Start with any feasible vector $b_1,\ldots, b_{\sf K}$ and consider  two indices $i< j$ with non-zero masses, $b_i>0$ and $b_j>0$. Updating this vector as  
\begin{IEEEeqnarray}{rCl}
b_i'=b_{i}-\Delta, \qquad &\textnormal{ and }& \qquad b_{i+1}' = b_{i+1}+\Delta,  \IEEEeqnarraynumspace\\
 b_{j-1}'=b_{j-1}+\Delta,  \qquad &\textnormal{ and }& \qquad b_{j
}' = b_j-\Delta,
\end{IEEEeqnarray}
 for any $\Delta \in[0, \min \{ b_i,b_j\}]$, results again in a feasible solution vector, which has smaller objective function due to the convexity of the coefficients $\{C_{i}^{(t)}\}$. 

Applying this argument iteratively, one can conclude that there must exist an optimal solution vector where all entries are zero except for two masses $b_{k}>0$ and $b_{k+1}\geq 0$. Since $\sum_{i=1}^{\sf K} i b_i \leq \sf r \sf N$, the index $k$ cannot exceed $\sf r$. By the decreasing monotonicity of the coefficients $C_{i}^{(t)}$, the optimal solution must then be to choose   $b_{\lfloor \sf r\rfloor}>0$ and $b_{\lfloor \sf r\rfloor+1}\geq 0$ and all other masses equal to 0. Since there is a unique such choice satisfying $\sum_{i=1}^{\sf K} i b_i \leq \sf r \sf N$ and $\sum_{i=1}^{\sf K}  b_i =  \sf N$, this concludes the proof. 
}

\end{document}